\documentclass[slac_one]{revtex4}
\usepackage{graphicx}
\usepackage{fancyhdr}
\begin{document}

\title{Search for contact interactions at HERA} 

%

\author{A. Raval (for the H1 and ZEUS collaborations)}
\affiliation{Penn State University, University Park, PA 16802, USA}

\begin{abstract}
The H1 and ZEUS collaborations at HERA have searched 
for signatures of physics beyond the Standard Model 
using high $Q^2$ neutral current deep inelastic 
electron-proton and positron-proton scattering events.
No significant deviations 
from Standard Model predictions were observed. 
Various $eeqq$ contact interaction models 
have been considered.
Limits on  the compositeness scale in general $eeqq$ contact 
interaction models,  mass to the Yukawa coupling ratio for heavy leptoquarks,
the effective Planck mass scale in models with large extra dimensions
and the effective quark charge radius are presented. 
\end{abstract}

\maketitle

\thispagestyle{fancy}

\section{INTRODUCTION} 
The H1 and ZEUS experiments at HERA (DESY, Hamburg) have facilitated the study 
of electron-proton and positron-proton collisions at center
of mass energies of up to 920 GeV.
During the HERA~I (1994-2000) running phase, about 100 $pb^{-1}$
of data were collected per experiment, most of it coming from $e^+ p$ 
collisions.
The collider was upgraded in 2000-2001, allowing for a significant
increase of luminosity. During the second phase of running (2002-2007, HERA~II), about 400 $pb^{-1}$ of data  per experiment were collected.
Additionally, spin rotators installed at the H1 and ZEUS interaction
regions provided longitudinal electron and positron polarization.
With an average lepton beam polarization of about 30-40\% and a significant
increase of integrated luminosity, particularly for the 
$e^- p$ sample, HERA~II has significantly increased the sensitivity 
of the experiments to physics beyond the SM.

\section{CONTACT INTERACTIONS}

New interactions between electrons and quarks involving mass scales 
above the center-of-mass energy can modify the  deep inelastic $e^\pm
p$ scattering cross sections at high $Q^2$ via virtual effects,
resulting in observable deviations from Standard Model (SM) predictions.
Four-fermion contact interactions form an effective theory, one which
allows us to describe such effects in the most general way.
Vector $eeqq$ contact interactions considered at HERA
can be represented 
as an additional term in the SM Lagrangian:
\begin{eqnarray*}
L_{CI} & = & \sum_{i,j=L,R} \eta^{eq}_{ij} (\bar{e}_{i} \gamma^{\mu} e_{i} )
              (\bar{q}_{j} \gamma_{\mu} q_{j}), 
\end{eqnarray*}
where the sum runs over electron and quark helicities,
and a set of couplings  $\eta^{eq}_{ij}$ describe 
the helicity and flavor structure of the contact interactions.
Hence various scenarios, with different chiral structures are possible.
Another model which may introduce a new interaction between electrons and quarks is one with large extra dimensions~\cite{add}. 
Lastly, an approach to probe quark substructure is by
 considering quark form-factors which depend on the effective quark radius.
The common feature of all these
 models is that they alter the cross-sections observed in neutral current
 deep-inelastic scattering (NC DIS) from that expected within the SM.
 Tests for such deviations have been performed and limits on the model parameters have been derived by H1 and ZEUS ~\cite{h1ci,zeusci,zeusci2}.

\section{RESULTS}

\begin{figure}[htb]
\centerline{
\includegraphics[width=0.55\columnwidth]{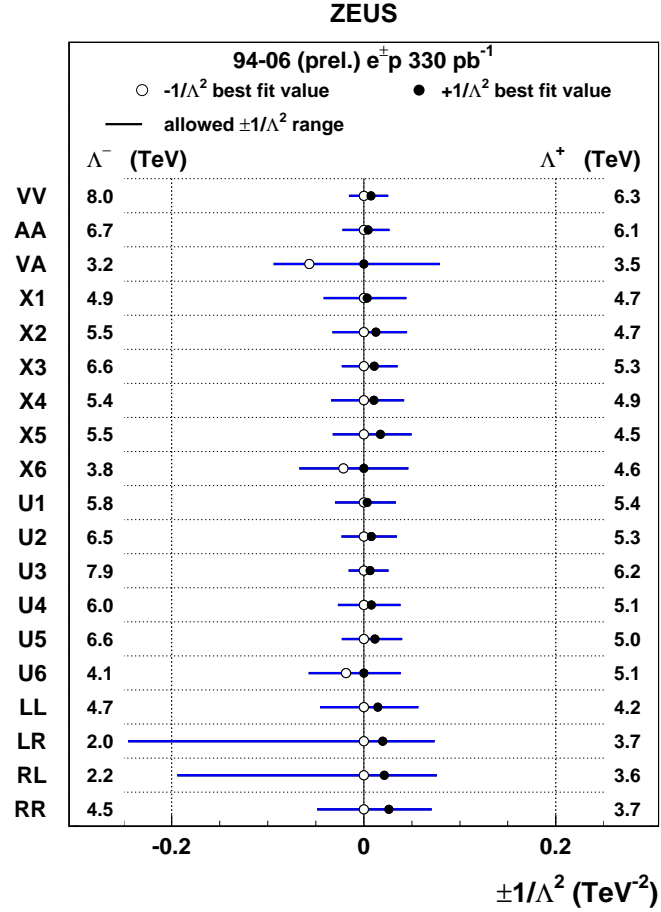}}
\caption{Results for general contact interaction models
  (compositeness models) obtained using the combined  $e^{+}p$ and $e^{-}p$ 
  data from ZEUS (1994-2006). Horizontal bars indicate the 95\% CL limits on
  $\eta / 4\pi = \varepsilon / \Lambda^{2}$; 
  values outside these regions are excluded.
  $\Lambda^{\pm}$ are the  95\% CL limits on the compositeness scale
  for $\varepsilon = \pm 1$.}\label{Fig:CI}
\end{figure}

For compositeness models, 
limits on the effective ``new physics'' mass (or compositeness) scale, $\Lambda$, 
are extracted assuming the relation 
$\eta = \pm 4\pi / \Lambda^{2}$.
Figure~\ref{Fig:CI} shows the results obtained by the ZEUS 
experiment for different compositeness models,
 based on the analysis of
1994-2006 data.
Limits on the effective mass scale $\Lambda$
range from 2.0 up to 8.0 TeV.
Corresponding limits obtained by the H1 collaboration, based
on the HERA I data only, range from 1.6 to 5.5 TeV~\cite{h1ci}.

\begin{figure}[p]
\centerline{
\includegraphics[width=0.57\columnwidth]{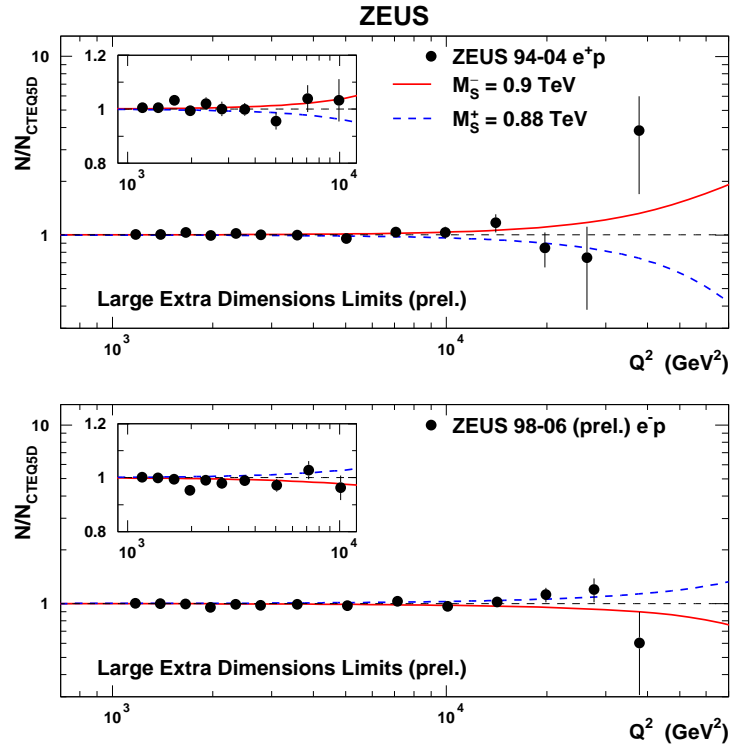}}
\caption{ZEUS data compared with 95\% CL exclusion limits for 
  the effective Planck mass scale in models with large extra 
  dimensions for positive ($M_{S}^{+}$) and negative  ($M_{S}^{-}$) 
  couplings. Results are normalized to the
  SM expectations using CTEQ5D parton distributions.}\label{Fig:LED}
\end{figure}

For the model with large extra dimensions~\cite{add}, both
collaborations have set limits on the effective Planck mass scale $M_{S}$.
For negative couplings, scales below
0.90~TeV (ZEUS 1994-2006~\cite{zeusci2}) and 0.78~TeV (H1 1994-2000~\cite{h1ci}) 
are excluded at 95\% CL. For positive couplings, the limits
are 0.88~TeV and 0.82~TeV, respectively.
Possible effects of graviton exchange on the $Q^{2}$ distribution
of NC DIS events, as measured by ZEUS, are shown in Figure~\ref{Fig:LED}.

 \begin{figure}[p]
\centerline{
\includegraphics[width=0.6\columnwidth]{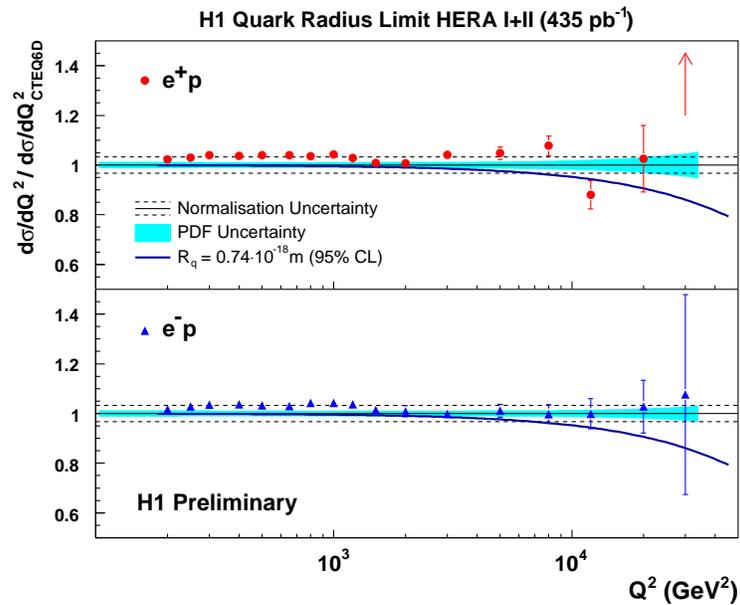}}
\caption{H1 data compared with 95\% CL exclusion limits for 
  the effective radius of the quark.
Results are normalized to the
  SM expectations using CTEQ6D parton distributions.}\label{Fig:Rq}
\end{figure}

 \begin{figure}[htb]
\centerline{
\includegraphics[width=0.44\columnwidth]{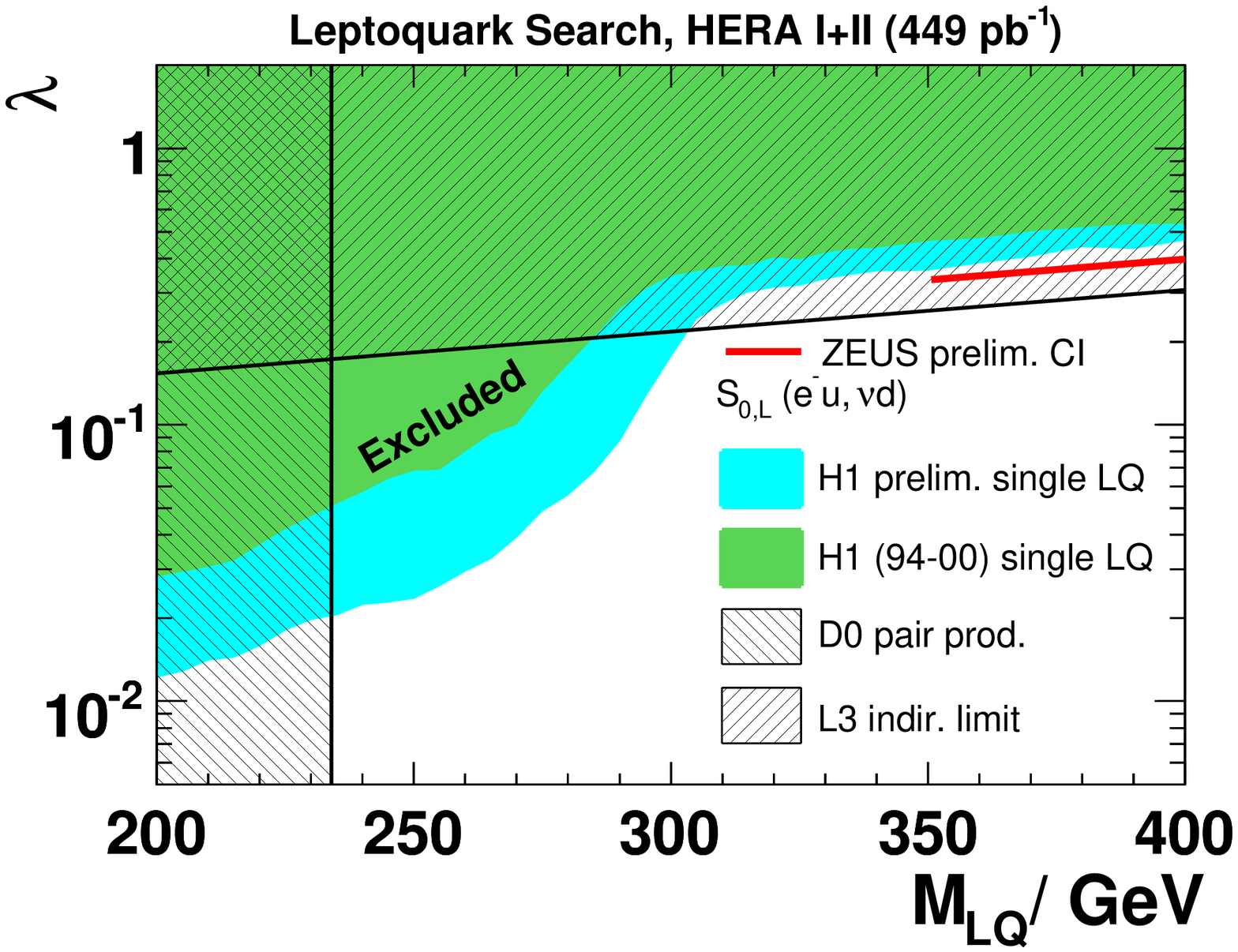}
\includegraphics[width=0.44\columnwidth]{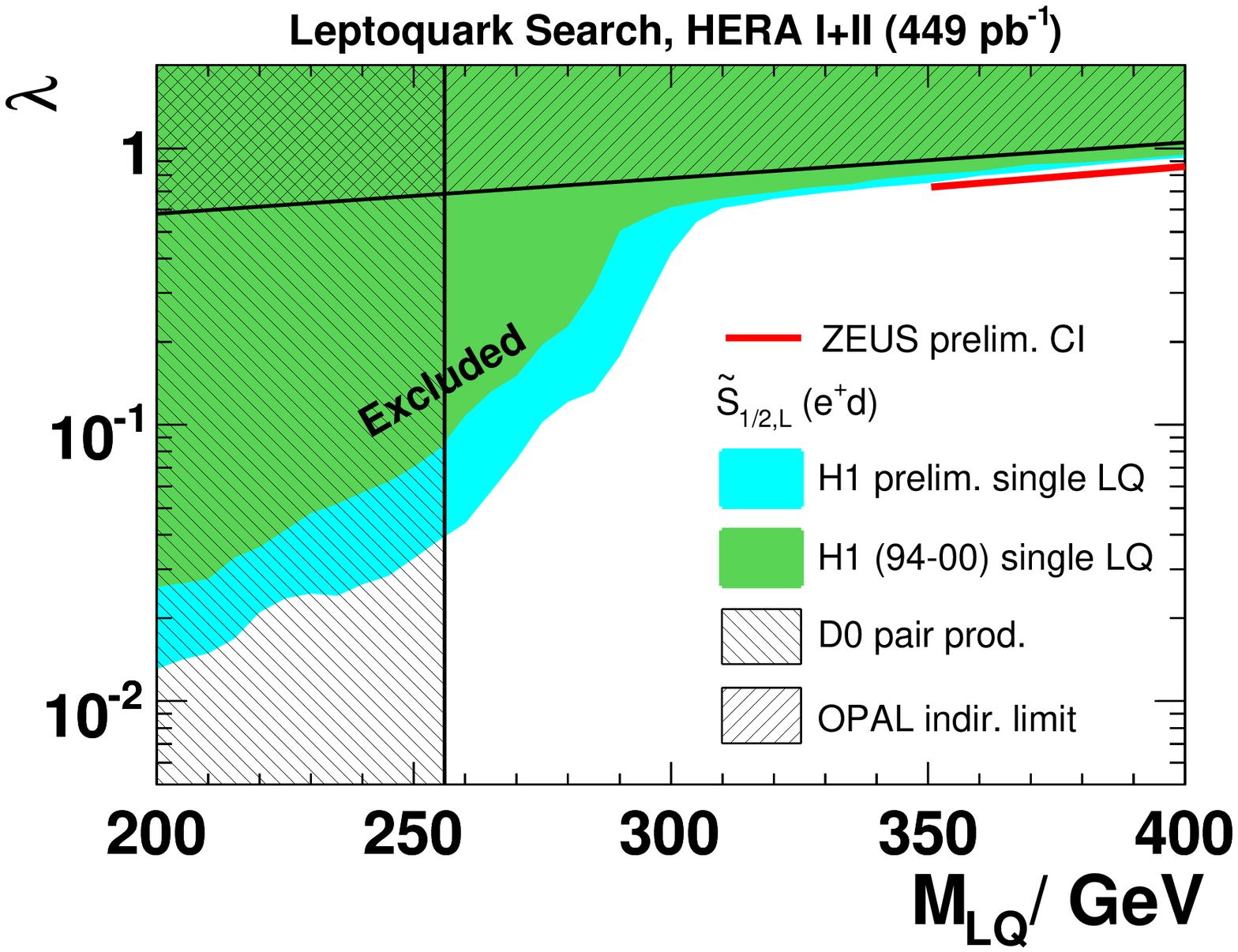}}
\caption{H1 exclusion limits at 95\% CL on the coupling 
 as a function of the leptoquark mass for $S_{0,L}$ 
and  $\tilde{S}_{1/2,L}$ leptoquarks~\cite{h1lq}. 
The indirect limits from ZEUS and L3 and the direct D0 limits are shown 
for comparison. }\label{Fig:LQ}
\end{figure}

Possible quark substructure can be searched for by  measuring
the spatial distribution of the quark charge.
Using the ``classical'' form factor approximation and assuming that
both the electron and exchanged bosons are point-like, limits on the
mean-square radius of the electroweak charge of the quark can be set.
From the analysis of the combined HERA I and HERA II  datasets, quark radii
larger than $0.74\cdot 10^{-16}$~cm (H1) 
and $0.62\cdot 10^{-16}$~cm (ZEUS) have been excluded at 95\% CL.
Figure~\ref{Fig:Rq} shows the H1 data with 
95\% CL exclusion limits for the effective 
radius of the quark~\cite{h1rq}.

In the limit of a
large leptoquark mass, $M_{LQ} \gg \sqrt{s}$, contact interactions can also be used to describe the effects of
virtual leptoquark production or exchange at HERA.
Utilizing the data taken between 1994 and 2006, the ZEUS collaboration has 
constrained
the leptoquark Yukawa coupling for different leptoquark types and masses.
Limits on the ratio of the leptoquark mass to Yukawa coupling range from 
0.29 to 2.08~TeV.
Figure~\ref{Fig:LQ} compares the ZEUS indirect limits to 
H1 exclusion limits (at 95\% CL) on the Yukawa coupling of $S_{0,L}$
and $\tilde{S}_{1/2,L}$ leptoquarks as functions 
of their masses.
Also shown are the limits from LEP and the Tevatron.

\section{CONCLUSIONS}

The high luminosity delivered at HERA in conjunction with
 lepton beam polarization
have opened up a new window for precise EW studies and searches for physics
beyond the Standard Model.
Measured NC DIS cross sections at high $Q^2$ are in very good 
agreement with the SM, hence limits on deviations from the SM could be set 
within several different models.
HERA running has finished, however analyses of large samples of data
continue. Updated results can therefore be expected in the near future.


\begin{footnotesize}

\end{footnotesize}


\end{document}